# Viral epidemics in a cell culture: novel high resolution data and their interpretation by a percolation theory based model

Balázs Gönci[1], Valéria Németh[1], Emeric Balogh[1,2,3], Bálint Szabó[1],
Ádám Dénes[4,5], Zsuzsanna Környei[4*] and Tamás. Vicsek[1,6]

[1]*Department of Biological Physics, Eötvös University,*
*Pázmány Péter s. 1A, H-1117 Budapest, Hungary*

[2]*Babes-Bolyai University, Department of Theoretical Physics,*
*Str. Mihail Kogalniceanu 1, RO-400084 Cluj-Napoca, Romania*

[3]*University of Szeged, Department of Optics and Quantum Electronics,*
*Dóm tér 9., H-6720, Szeged, Hungary*

[4]*Institute of Experimental Medicine of the Hungarian Academy of Sciences,*
*Szigony u. 43., H-1083, Budapest, Hungary*

[5]*Faculty of Life Sciences, University of Manchester,*
*A. V. Hill Building, Oxford Road, Manchester, UK*

[6]*Statistical and Biological Physics Research Group of the Hungarian Academy of Sciences, Pázmány Péter s. 1A, H-1117 Budapest, Hungary*

*Corresponding author:

Tamás Vicsek

*Eötvös Loránd University (ELTE)*
Department of Biological Physics
H-1117 Budapest,
Pazmany P. Stny 1A, Room 3.75
HUNGARY

Phone:
(+36-1) 372-2755

e-mail: vicsek@hal.elte.hu   home: http://angel.elte.hu/~vicsek

**Abstract**

Because of its relevance to everyday life, the spreading of viral infections has been of central interest in a variety of scientific communities involved in fighting, preventing and theoretically interpreting epidemic processes. Recent large scale observations have resulted in major discoveries concerning the overall features of the spreading process in systems with highly mobile susceptible units, but virtually no data are available about observations of infection spreading for a very large number of immobile units. Here we present the first detailed quantitative documentation of percolation-type viral epidemics in a highly reproducible *in vitro* system consisting of tens of thousands of virtually motionless cells. We use a confluent astroglial monolayer in a Petri dish and induce productive infection in a limited number of cells with a genetically modified herpesvirus strain. This approach allows extreme high resolution tracking of the spatio-temporal development of the epidemic. We show that a simple model is capable of reproducing the basic features of our observations, i.e., the observed behaviour is likely to be applicable to many different kinds of systems. Statistical physics inspired approaches to our data, such as fractal dimension of the infected clusters as well as their size distribution, seem to fit into a percolation theory based interpretation. We suggest that our observations may be used to model epidemics in more complex systems, which are difficult to study in isolation.

INTRODUCTION

Epidemic is a common phenomenon in a wide variety of systems ranging from flora through domestic animals to human populations. Understanding the statistics of the dynamics and spatial aspects of spreading infection could be crucial in its containment [1-4]. Appropriate predictions about epidemics of various kinds can lead to enormous savings of both material goods as well as human lives [5-8]. In spite of its unquestionable importance, relatively little is known about the details of the spreading of a disease in a natural environment. In particular, observations of epidemics on a large scale (involving tens of thousands of organisms) are extremely scarce and incomplete [1, 2, 9, 10]. There are obvious reasons for this: i) in the case of individuals having either a considerable size (e.g., trees or animals), or moving around extensively (e.g., birds or people), the area to be covered in a controlled experiment is huge. ii) In addition, to infect and study living subjects is a risky and ethically problematic procedure. Therefore, a majority of the related studies involves modelling [3, 11] instead of experimenting. Many studies have pointed out the relevance of spatial inhomogeneities, to the specific (e.g., scale-free) patterns of connectivity and transport of the subjects. However, in the case of static units [12, 13] the possibility of a detailed comparison between the predictions of the models and the actual observations for situations involving inhomogeneities has remained limited.

In general, there are two major types of infection spreading: i) the position of the infected units does not change significantly in time, e.g., in cells in a tissue exposed to viral or bacterial infection [14] or trees in a forest [15] ii) the organisms carrying the infection are moving around and thus are acquiring new sets of connections (this is perhaps the more common case, in the case of people, e.g., involving long distance flights, etc.). The latter case has two variants [5], the units only diffusing around, or, in the case of birds or people, they can make huge distances in a short time so that the spatial aspect is less relevant [6-8] and a network description becomes more appropriate [16, 17]. The former case applies to serious diseases caused by microorganisms such as Burkholderia spp., Listeria monocytogenes [18], Mycobacterium marinum [19], Shigella spp. [20], and Rickettsia spp [21]. These bacterial pathogens enter host cells and spread through tissues by moving directly from one cell into adjacent cells using an actin polymerization driven tail propelling them inside the cytoplasm. The *in vitro* spreading of Listeria [14] and Shigella [22] in 2D cell cultures was shown to lead to circular or more complex plaques. Finally, a standard paradigm used to explain infection spreading through a percolation-like process involves a forest in which some of the trees become infected by a pathogen that can spread to a neighboring tree only with a probability smaller than 1.

In order to handle the above mentioned constraints we have designed experiments allowing us to study in detail the outbreak dynamics and the patterns of infection spreading in a system containing tens of thousand of essentially static units. The progression of the epidemics was followed by both making video recordings and high resolution photos (i.e., tracking the infection by both GFP expression and making preparations with various selective, immuno-histochemistry based stainings). With the help of a computerized system at each given time 3 x 3 = 9, 4 x 4 = 16 (movies) and 10 x 10 = 100 (microscopic fields) images were stitched together to obtain uniquely high resolution (e.g., the merged pictures of the stained plates have appr. 13000 x 9000, i.e., around 100 million pixels.).

To interpret our observations we use models, quantities and expressions primarily associated with percolation theory [23]. This theory predicts that for the simplest version of a growing percolation process (in which occupied sites are randomly added to an initially empty set of lattice sites), close to the transition (or critical) point, when the size of the largest cluster becomes compatible with the system size, the distribution of the cluster sizes follows a power law with corrections. In particular, if $n_s$ denotes the normalized number of clusters as a function of their size s than

$$n_s(p-p_c) \propto s^{-\tau} f(s|p-p_c|^{-\sigma}),$$

where p is the proportion of occupied sites, $p_c$ is the critical proportion (or probability of being occupied), $\sigma$ is a positive number and f(x) is a scaling function being about constant for x<<1 and exhibiting a fast (exponential) decay for large values of its argument. Thus, close to the transition point (p is about $p_c$) $n_s$ is expected to exhibit as a power law decay with an exponent $-\tau$ followed by an exponentially fast cut-off depending on how far the system is from its critical point. We test the validity of this and other related predictions of the theory and build and simulate a model which is in agreement with the experiments.

In short, i) our study is aimed at exploring a common and important but yet not fully documented phenomenon, ii) we use an image processing technology far beyond those used in the present context before and iii) we successfully interpret our findings using numerical simulations.

RESULTS

Our experimental setup (**Fig.1.**) was based on a tissue cell culture (confluent monolayer of astrocytes) exposed uniformly to a recombinant pseudorabies virus (PRV) Ba-DupGreen (BDG), which expresses green fluorescent protein (GFP) in the infected cells with immediate early kinetics. We have chosen astrocytes from a number of cell types examined in preliminary experiments, as these cells show restricted release of infectious viral particles whereas the spread of infection among interconnected cells is not compromised [24].

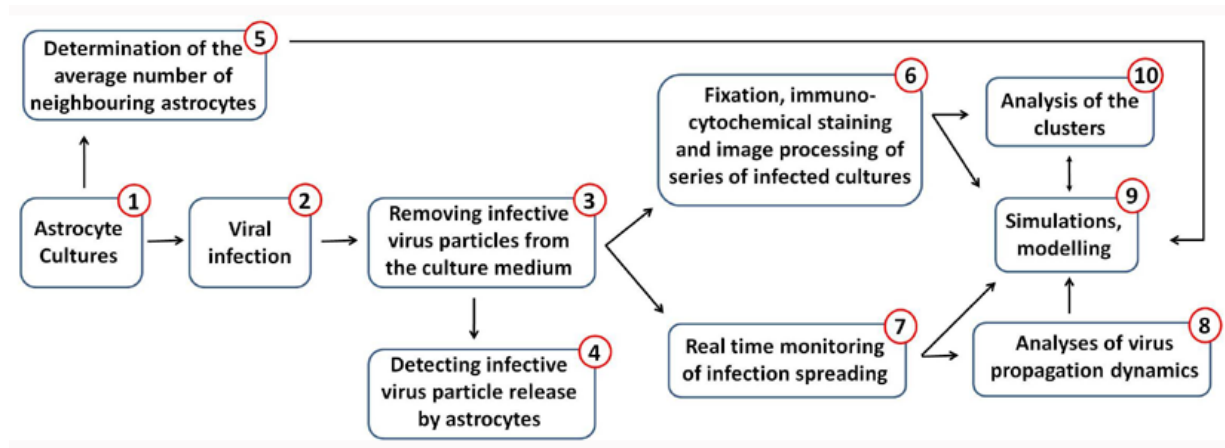

*Figure 1. Flow diagram of the experimental design. All steps are discussed in details throughout the text as well as in Materials and Methods.*

First, we initiated productive infection provoking GFP expression in a limited number of cells (see **Materials and Methods**). After removing the infective viral particles from the culture medium (**Table S1**) we obtained an extensive series of images of the developing epidemics (growing clusters of infected cells; **Fig. 2.**). These images, especially after computer aided processing and analysis revealed a set of highly complex geometrical patterns.

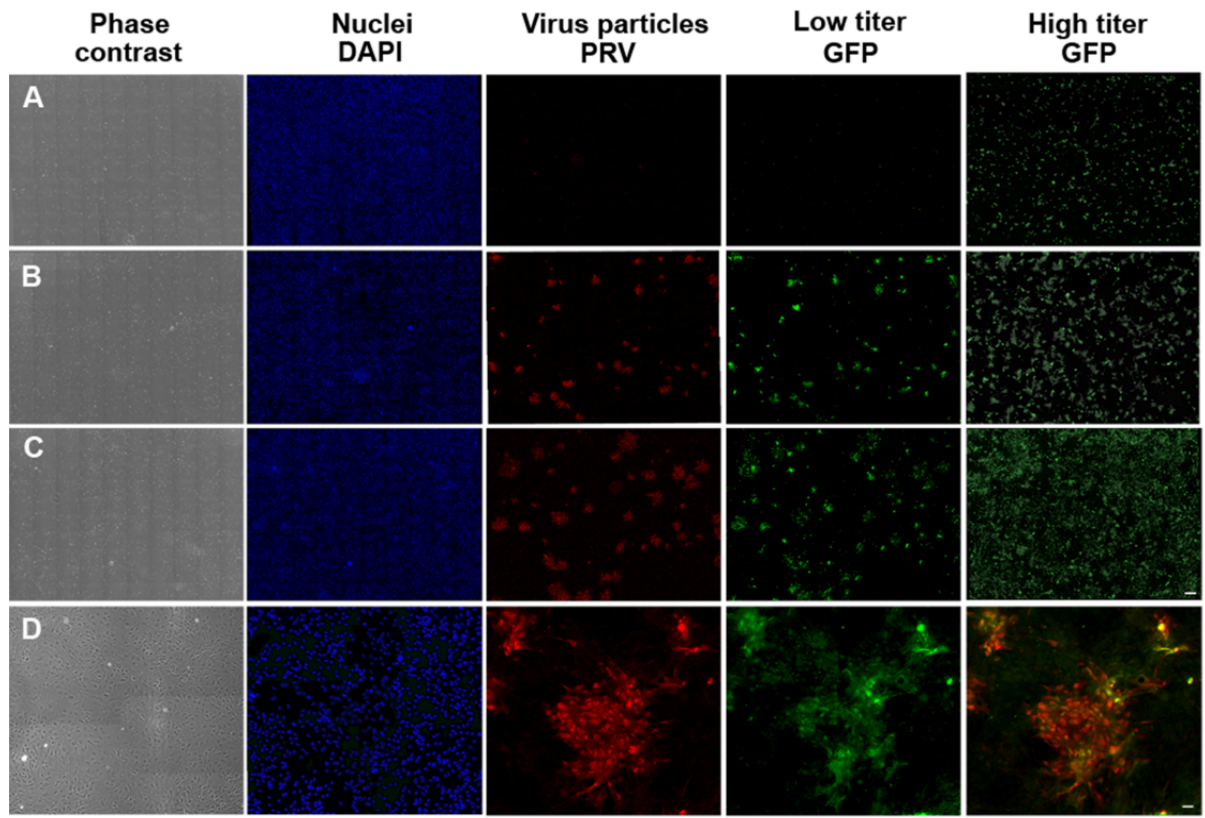

*Figure 2. Mosaic images of astrocytic cultures. The cultures were fixed 18 h (A), 30 h (B) and 42 hours (C) after infection with the BDG-PRV virus. In the bottom row (D) higher magnification images can be seen taken 36 hours after infection. The last image shows accurate overlap between immunostained PRV virus particles (red) and GFP (green) expression by infected cells. Scale bars are 300 μm and 25 μm, respectively. High resolution file can be downloaded from: http://amur.elte.hu/BDGVirus/*

In **Fig. 3.** we displayed a representative example showing the rich spatial structure of the infected zones close to a point when the largest cluster spans the whole culture. An approximate self-similarity could be observed being visually demonstrated in the figure by including consecutively enlarged parts of the original image (supporting information **Fig. S1**).

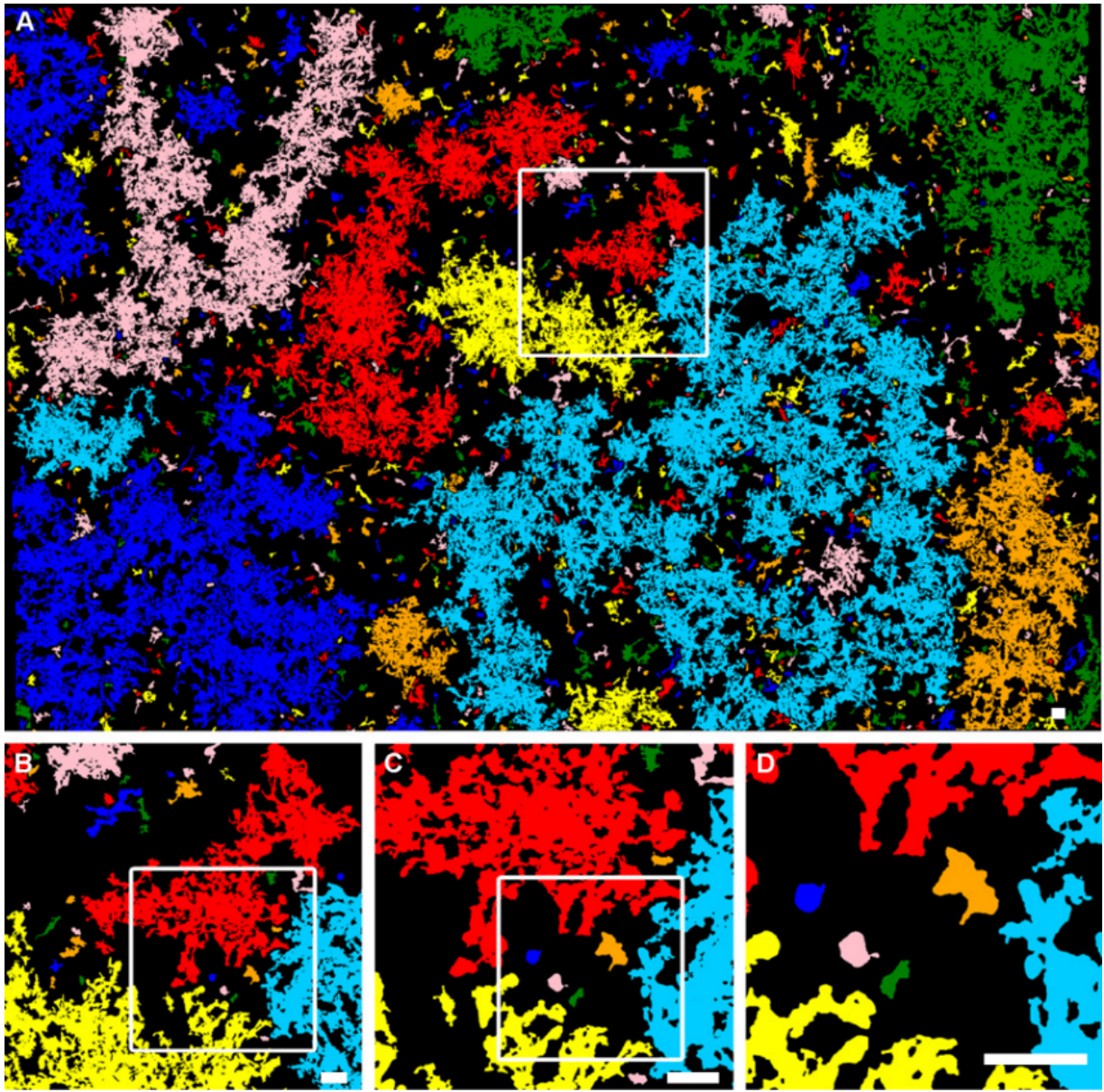

***Figure 3.*** *An image processed representation of a large infected region of the monolayer of astrocytes. 100 (10 x 10) microscopic field images were taken from fixed cultures with GFP expressing infected cells and stitched together to obtain a high resolution single picture (approximately having 13,000 x 9,000 pixels). Then, the originally black and white image was analysed by finding clusters of cells (connected regions of larger than 1000 pixels: which number was associated with the average cell size). Next, each separate cluster was coloured by choosing a colour from a random set of 8 possible ones. The obtained clusters and their perimeters display a highly complex spatial structure reminiscent of mathematical fractals and possessing some level of self-similarity illustrated by sequential magnification of parts of the structure. Scale bar: 100 μm. (See original image in **Fig. S1**.)*

Since the images we obtained were reminiscent of those generated by simple statistical physics models of percolation, we determined both the distribution of the cluster sizes and the fractal dimension of the largest clusters (**Fig. 4. and Fig. 5.**, respectively). These two quantities are considered to be the most characteristic features of percolating systems and are known to be universal over a wide selection of model variants.

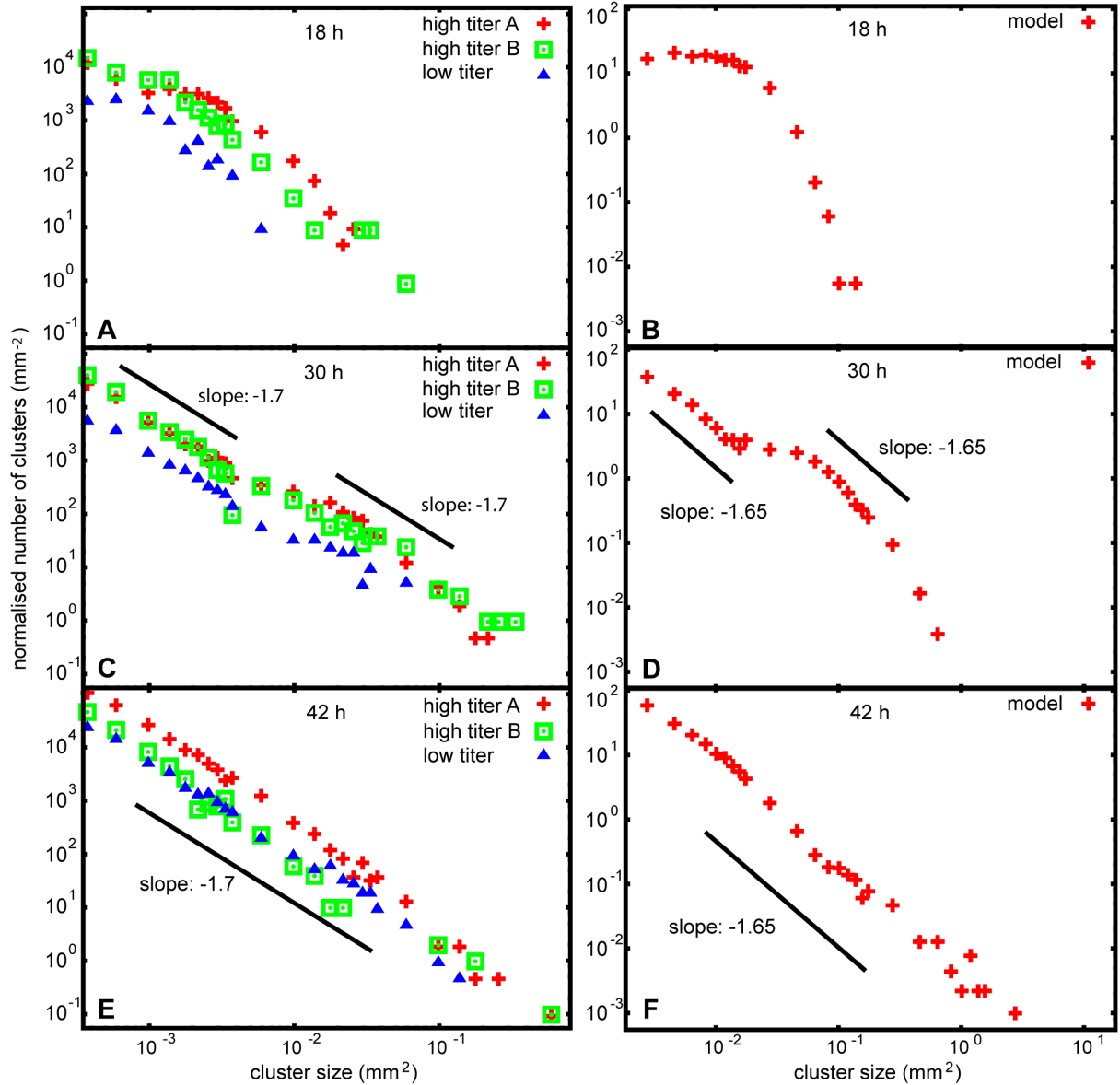

*Figure 4. The distribution of cluster sizes. Panels **A**, **C** and **E** show the cluster size distribution after 18, 30 and 42 hours, respectively, as determined from a series of pictures analogous to **Fig. S1**. Corresponding cluster size distribution of the model is presented in panels **B**, **D** and **F**. Experimental graphs are consistent with the predictions of our percolation based simple spreading model: both the experimental scaling behaviour at 42 hours and the spreading of the infection before that are well characterized by the model. A scaling (power law) distribution is built up in time, i.e., the number of infected clusters vs cluster size follows a power function instead of an exponential drop. Model time was scaled to fit to the infection spreading rate of high titer experiments.*

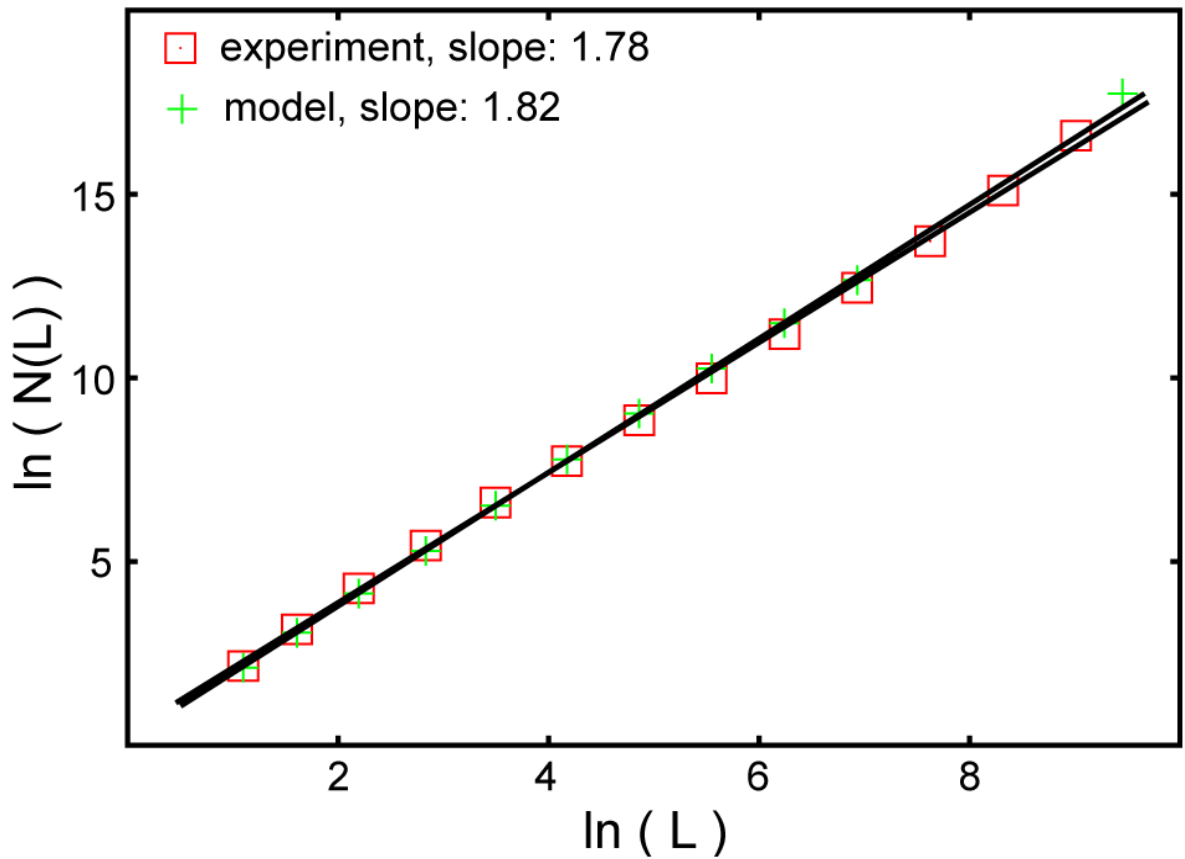

*Figure 5. Determination of the fractal dimension of the largest clusters. This figure (squares) shows the number of cells in the largest clusters of infected cells as a function of the linear size (L) of the regions in which they were counted. The fitted line indicates that the geometry of the largest infected regions is analogous to those of fractals with a dimension of about 1.78. Remarkably, the simulation data (crosses) are almost indistinguishable from those obtained from the observations.*

Indeed, our experimental images also displayed features very much like percolation models, i.e., the largest cluster had a well defined fractal dimension and the cluster size distribution function could be interpreted as an algebraically decaying function with some characteristic modifications (**Fig. 6.**).

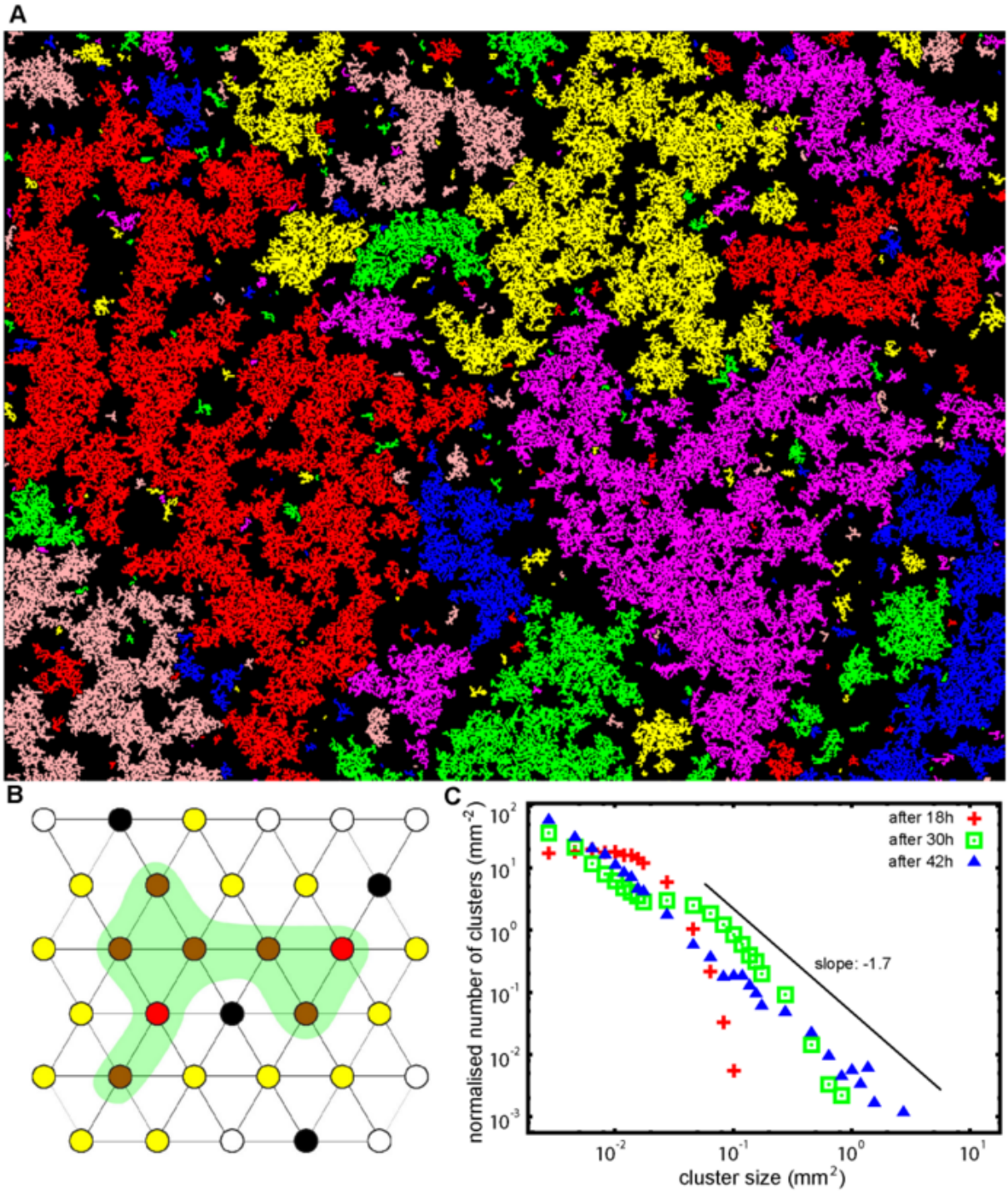

*Figure 6. Results from the simulations of our model. A representative set of clusters (bottom 20% of the original picture has been omitted to save space) is displayed (top, **A**) for the following parameters: size 500 x 500 lattice units, 500 starting seed particles with a "switching on" time distribution following a lognormal fitted to the experimentally observed parameters (mean=7.5 h, standard deviation=2.5 h) plus 2000 further seeds with a constantly increasing rate in time between 20 h and 42 h (corresponding to the observed second wave of infection, see **Fig. S4.**). The implemented values of the two model parameters were ρ=0.0182 and α=3. The bottom left (**B**) figure is a schematic demonstration of the model where the coloured circles correspond to the experiment as follows: black dots → not susceptible or missing cells; red dots → primarily infected cells; brown dots → cells infected by one of their neighbours; yellow dots → cells potentially infected in the next time step;*

*green hollow: a cluster of infected cells. The bottom right (**C**) plot demonstrates that the cluster size distribution in the model also displays a scaling behaviour for the parameters corresponding to the experimental conditions (the same as for the picture in the top).*

Associating the infected cells with the occupied sites, we saw that the observations are rather reminiscent of the extremely simple percolation model. There were some differences, however, to be discussed later. The value of approximately 1.7 we found for the exponent $\tau$ (**Fig. 4.**) defined in the introduction is consistent with both the value obtained from our simulations (approximately 1.65) and with the predictions for the distribution of cluster sizes in systems under a non-equilibrium, cluster growth regime (in equilibrium systems $\tau > 2$) [23].

In addition to static configurations, we have also investigated the dynamics of the infection spreading in our tissue cultures (**Fig. 7.** and supplementary **Videos S1-S6.**). Long term video microscopy data have enabled us to keep track of the number of infected cells after all of them had been exposed to a uniformly distributed infection for a short (as compared to the experiments' time scale) time interval. The Supplementary **Video S5** clearly shows that the individual astrocytes witin the monolayers do not shift from their original position more, than a fraction of their diameter, within a considerable amount of time (more than 24h).

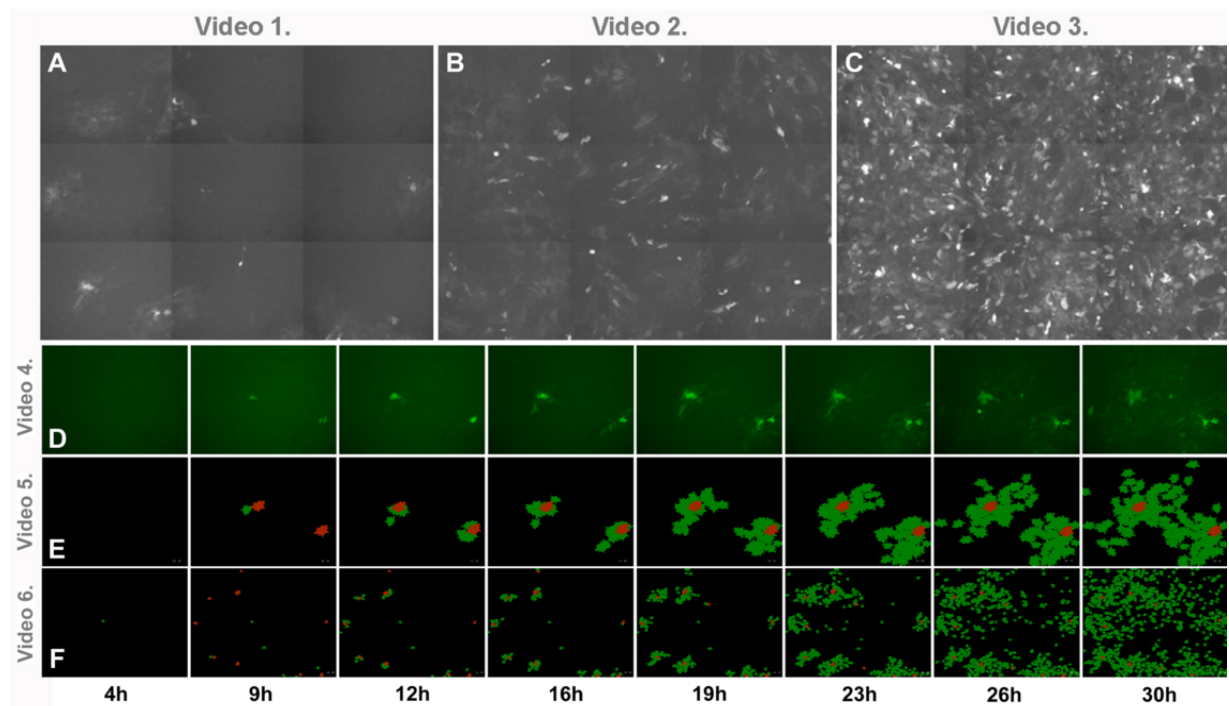

*Figure 7. Images demonstrating the spreading of the infection. Screenshots (**A-D**) and graphical representations (**E,F**) of videos on virus propagation within astroglial cultures treated with BDG-PRV virus. Supplementary **Videos 1., 2.** and **3.** were recorded using cultures infected with different titers showing virus spread on large scale. Titers in these experiments were **A**) 2,5x10$^4$ PFU/ml ("low" titer) **B**) 2.5x10$^5$ PFU/ml ("high" titer) **C**) 1,25x10$^6$ PFU/ml ("highest" titer). Image series **D**) and **E**) and supplementary **Videos 4. and 5.** show the appearance of GFP in "low titer" infected cells over time within a representative microscopic field. Formation of cell clusters with more advanced infection can be seen in image series **F**) and **Video 6.**, representing 9 stiched microscopic fields. Red cells indicate*

the foci of the developing clusters. High resolution videos and images can be downloaded from: http://amur.elte.hu/BDGVirus/

We have determined the rate by which cells acquire a "primary" infection (directly related to the initial exposure), and also the dynamics by which the number of infected cells grows as a result of "secondary" infections (infection due to being a neighbour of an already infectious cell; **Table S2**, **Fig. S2)**. The motion picture recordings allowed us to gather some unique statistical data about the time development aspects of the phenomenon as well (**Fig. 8.**).

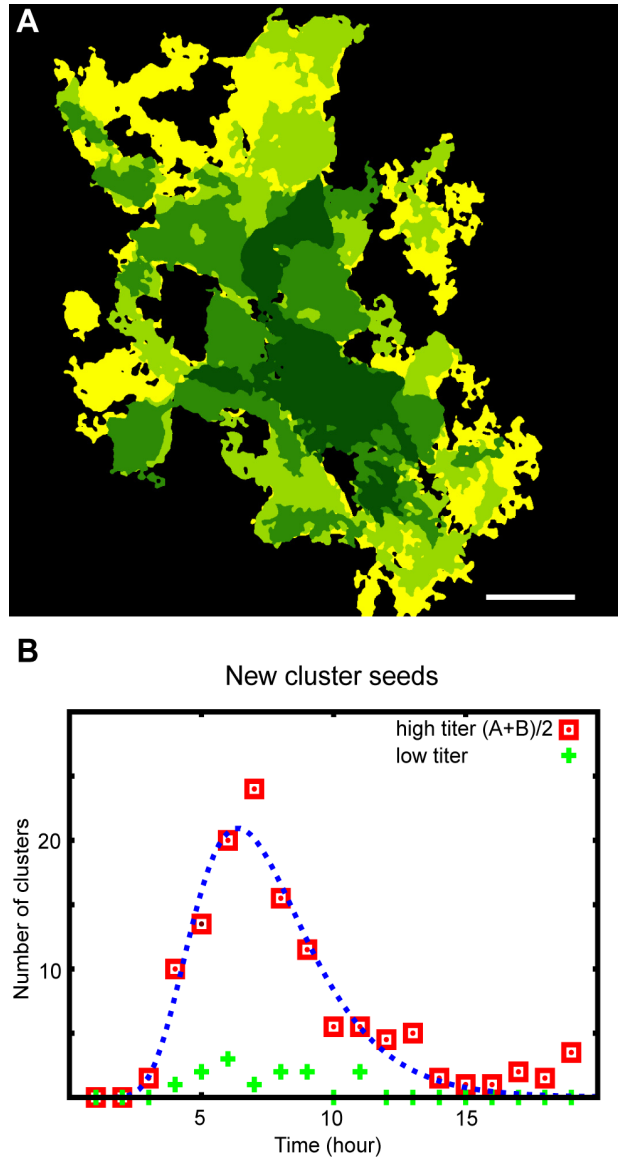

*Figure 8.* *Visualization of the spreading of the infection around a single infected seed. The upper picture (**A**) demonstrates the progression of the infection around a single infected cell (see **Videos 4-6.**). The colour coding is such that lighter colours correspond to later times and the movie frames (also image processed here) were taken at time 15, 18, 21, 24 hours after the initial exposure to the virus. The bottom plot (**B**) shows the probability distribution function for a single isolated cell (playing the role of a seed for the infection spreading) to become infected. Fitted is a log-normal function giving a significantly better agreement than a Gaussian. Scale bar: 50 μm.*

Next, in order to both qualitatively and quantitatively interpret our large scale spatio-temporal epidemiological data, we have designed a model. This approach enabled us to estimate to what degree the global behaviour is sensitive to details of the processes taking place on the local (cellular) level. Our model was, on one hand based on the simplest prior epidemic models (percolation), on the other hand it took into account some of the specifics of the present experiment in a new way. The two major features of the model were: i) neighbours (of the already infected cells) were infected with a probability taken from an algebraic (non-linear) distribution, ii) a secondary wave of infection was represented by the gradual introduction of new infectious sites (see **Table S2** and MATERIALS AND METHODS for details). These simulations allowed us to calculate the same quantities which we have determined for the experimental images and thus, a quantitative comparison could be made concerning the essential parameters of the epidemics.

We expected that a cluster grown from a single seed would have had a well defined fractal dimension [25]. However in our case there were many clusters in the system growing simultaneously and sometimes these clusters merged. This second mechanism would have changed the fractal dimension of the formed clusters considerably. The merging of the clusters could be observed also in the experiment. The simulation was stopped when the ratio of the infected cells reaches a given value calculated from the experimental results. After this point the infected cells began to die. A typical set of clusters obtained for parameters corresponding to the experiments is shown in **Fig. 6.** with the different clusters having selected colours by size from a table of eight. The dynamics of cluster growth of the model can be seen in supplementary **Video S7.**

DISCUSSION

Our infection spreading experiments have been possible because the viral particles are able to spread within the compromised cell and enter its processes. Newly formed virions have the potential to infect nearby cells following exocytosis from the infected cell and a subsequent uptake through the cell membrane. In the case of productive infection, virus containing cells produce sufficient progeny virions to infect their surroundings and to initiate a focal spread of infection over time. Compromised (but viable) as well as dying cells are also capable of releasing infectious virus particles into the media, which may attach to and infect further cells from a large distance to the place of the original infection [26-31].

Detailed documentation of the spreading of viral infection in a layer of many thousands of cells has been carried out in a very limited number of papers only [32, 33]. These prior investigations involved systems in which an infected cell was likely to infect all of its neighbours in a time scale much shorter then that of the whole experiment. In addition, the infected cells introduced back to the medium significant amounts of viral particles. Under these conditions the infected regions of the culture possess interesting ring-like or elongated shapes, allowing quantitative analysis of some of the features of the propagating zone of infection. We observed such convection current-induced disturbance of the infection spread in the case of some cell types used in preliminary experiments, such as swine testis cells (ST) or canine (MDCK) and human kidney cells (HEK293) (data not shown). The phenomena we observed for the above cell types were very much like those described in refences [34-36]. In the case of the PRV virus versus astrocytes system, however, neither the infection of all of the neighbours nor the secondary infection of the medium played a dominant role and, in fact, this is why we were able to observe the growth of percolation-cluster-like infected regions. In contrast to the cell lines mentioned above, which - due to rapid cell lysis - released a very high titer of infectious viral particles into the culture medium (data not shown), low virus titers in the supernatant of infected astrocytes did not lead to convection current-induced disturbance of the infection spread and, in line with simulation data, the growth of newly formed clusters after 18 hours of infection did not have a relevant impact on the percolation-type growth. Another possible advantage of using primary astrocytes over cell lines may be that due to cell to cell contact inhibition, the proliferation of primary astrocytes is much slower than that of cell lines. This made possible to image these cultures even beyond 48 hours, without the disturbing effects of an over-confluent multilayer, which would also involve significant cell death.

Several biological and physical parameters are known to affect the cells` susceptibility to infection and/or lead to inhomogeneous spreading of infection in both *in vivo* and cell cultures. Variability in cell cycle (both in cell lines and primary cells), expression levels and availability of different virus receptors, the variable activation state of individual cells, changes in fluid flow and diffusion can all contribute to changes in spreading characteristics [37-41]. In addition, in our case the actual geometry of the astrocytes the varying width and extension of their processes are likely to result in a highly variable number of contacts (through which the infection can take place) between them. The non-linear term (power law) we use in our model can account for such type of details because it corresponds to a situation, in which the probability of becoming infected in the next time step by a neighbour is not scattered around a well defined average but is very unevenly distributed.

Based on our experimental observations and the above description the following scenario can be proposed for the large scale spreading in a situation when i) the susceptible units are spatially fixed ii) the process of spreading is highly inhomogeneous. The inhomogeneities are likely to be due to the highly variable geometry of the individual cells as well as their log-normally distributed susceptibility to become infected after a sudden initial exposure to infection. We find that the actual spreading is a combination of the growth of individual plaques around their corresponding "seed" (initially infected cell) and the subsequent fusion of these plaques. However, the plaques themselves have a complex geometry since the infection rate of a cell is highly varying from cell to cell. In such situations, many small and a fewer (but not exponentially decaying number of) larger connected infected regions are expected to appear in the system. All this can be interpreted in terms of simultaneously growing percolation clusters. The model we proposed is extremely simple and is designed to account for the above experimental details. Both the visual appearance and the quantitative statistical features of the simulated clusters agree reasonably well with the observations.

Since our model possesses general features of live systems and takes into account the specificities of the currently used astroglial cell culture to a minimal degree only, we are motivated to suppose that our observations are likely to be applicable to a very wide variety of experimental situations being dominated by the infection taking place at spatially scattered points and spreading to neighbours with a rate highly varying from one susceptible units to another. In such cases the largest infected regions are likely to have fractal geometry and the size distribution of the clusters follow a scaling law. Such features have long been suggested by population dynamics models but here we present the first large scale experimental verification of the conjectured behaviours.

An interesting extension of the present work would be to consider cell populations in which the migration of the infectious cells cannot be ignored. In principle, even the network aspects of infection spreading could be studied in a similar setup with cells having very long processes capable of reaching out to and through them eventually infect distance cells.

MATERIALS AND METHODS

**Ethics Statement**.

This study was carried out in strict accordance with 86/609/EEC/2 Directives of European Community. Based on the permissions issued by the Food and Animal Health Authority, Budapest, Hungary, under the licenses No.: 2296/003/2006-2011 and 3668/003/2008-2014, the Animal Care and Experimentation Committee of the Institute of Experimental Medicine, Hungarian Academy of Sciences approves investigations, such as ours, complying with the above licenses during their period of validity.

**Materials.**

Trypsin, DNase I, poly-L-lysine, MEM, foetal calf serum, glutamine, Hoechst 33342 nuclear dye and anti-GFP antibody were purchased from Sigma-Aldrich Kft. Hungary; tissue culture dishes either from Greiner Bio One Co., Hungary or IBIDI GmbH, Germany. CellMask™ Orange, anti-rabbit-Alexa594 and anti-mouse-Alexa488 fluorescent dyes were obtained from Invitrogen Co., Hungary. Primary antibody to PRV virion proteins was a kind gift of Prof. Lynn Enquist.

**Experimental setup.**

The Methods section follows the Experimental setup explained in the Flow diagram of **Fig.1.**

**Astroglial cultures.**

Astrocytes were isolated from whole brains of neonatal (P0-P3) mice, as described earlier [42]. In brief, meninges were removed and the tissue pieces were subjected to enzymatic dissociation, using 0.05 % w/v trypsin and 0.05 % w/v DNase for 10 minutes at room temperature. The cells were plated onto poly-L-lysine coated plastic surfaces and were grown in Minimal Essential Medium supplemented with 10 % foetal calf serum, 4 mM glutamine and 40 μg/ml gentamycin in humidified air atmosphere containing 5 % $CO_2$, at 37 °C. The culture medium was changed on the first two days and every second or third day afterwards. Confluent primary cultures were harvested by trypsinization and re-plated onto pLL coated glass coverslips or into petri dishes, according to the actual experimental design. Secondary astrocytic cultures reaching confluency and displaying mosaic-like pattern were infected with PRV viruses, as it follows.

**Viral infection.**

We aimed to choose a virus strain for the experiments, which has relatively slow and well-controlled spreading characteristic in order to analyse the evolution of plaque size and distribution very precisely over time. To this end, Ba-DupGreen (BDG), a recombinant PRV-Bartha-derived pseudorabies virus was used. The virus genome was altered by insertion of the green fluorescent protein (GFP) gene expression cassette to the putative antisense promoter (ASP) located at the inverted repeat region of the virus [43]. Due to this modification, BDG multiplicates with a significantly lower rate *in vitro* and infects fewer numbers of cells *in vivo* than the parent PRV-Ba strain [44]. The reporter gene was placed under the control of the human cytomegalovirus major immediate early (IE)1 promoter (CMVP). Therefore, GFP was expressed at a very early stage of infection in the compromised cells that made possible to precisely follow the spread of infection over time. The detailed explanation of the mechanisms underlying the gene expression profile and the spreading kinetics of the virus was described previously [43,44]. The virus was grown in porcine kidney PK-15 cells to a titer of $6 \times 10^8$ plaque-forming units (PFU)/ml and was further concentrated to $1.5 \times 10^{9.8}$ PFU/ml by ultracentrifugation at 70,000×*g*. Virus stocks were aliquoted and stored at − 80 °C. Aliquots were thawed immediately prior to infection of astrocyte cultures. During preliminary experiments, the appropriate virus titer to obtain a low number of plaques was determined. The astroglia cultures were infected with BDG at a final titer of $2.5 \times 10^5$ PFU/ml ("high" titer) or $2,5 \times 10^4$ PFU/ml ("low" titer). The cells, grown in 35 mm diameter Petri dishes were incubated with 1 ml of the virus containing medium, for 1h, at room temperature. The multiplicity of infection (MOI) was ~0,17 or 0,017 PFU/cell, respectively.

**Removal of infective viral particles from the culture medium.**

In order to remove infective virus particles from the culture medium we washed all the cultures at least three times after the 1h viral exposure. To check whether three changes of medium was sufficient to completely remove viruses from the supernatant we transferred the first, second and third washes to intact cultures. 24 hours after either a 1 hour or a day long incubation period we fixed the cells and counted the GFP expressing plaques or cells within 30-30 random microscopic fields in each dish. Both the size of the plaques and the number of infected cells or cell groups was strongly reduced due to the washing procedure. The data indicated that 3 washes were sufficient to remove the "original" viral particles from the infected cultures (see supplementary **Table S1.**).

**Release of infective viral particles to the culture medium by astrocytes.**

To determine, whether the virus-treated cells released infective virus particles into the medium during culture maintenance we analyzed the infective capacity of the culture media removed after 1, 5, 24, 30 or 48 hour long incubation periods. Equal volumes of the media were transferred to non-treated sister cultures, which were further incubated for either 1 or 24 hours. Finally, we fixed the cells and counted the GFP expressing plaques or cells within 30-30 random microscopic fields on each coverslip (duplicates). The data indicate that ~24 hrs after the infection astrocytes began to release infective viral particles to the culture medium (**Table S2.**)

**Determination of the number of neighbouring astrocytes.**

To determine the average number of neighbouring astrocytes we stained the cultures both with CellMask™ Orange, a fluorescent plasma membrane stain of living cells, and Hoechst-33342, a membrane-permeable nuclear dye. With the help of the visualized cell contours and the nuclear stainings we determined the average number of the neighbours of any individual astroglial cell (n=100) (**Fig. S3.**). These data were further used in the development of the virus propagation model.

**Immunocytochemistry and image processing of fixed cultures.**

We used a Zeiss Axiovert 200M microscope equipped with AxioCamMrm to make images of astrocytic cultures fixed at various moments (3, 6, 9, 12, 18, 24, 30, 36, 42, 48, 72 hours) after infection. For detailed analyses we used 2-2 parallel cultures infected either with low or high virus titers. Immunocytochemical stainings were performed as described earlier [42]. In the present work, primary antibodies to structural virus proteins (PRV, rabbit, kind gift of Prof. Lynn Enquist) and to GFP (mouse) were used in 1:1000 and 1:2000 dilutions, respectively. As secondary antibodies anti-rabbit-Alexa594 and anti-mouse-Alexa488 fluorescent dyes were used in a dilution of 1:1000. DAPI was used to label cell nuclei. The stainings revealed that GFP expression within the infected cells overlaps with the PRV immunostaining. From each culture we made 10*10 overlapping images in four channels (DAPI, GFP, PRV and Phase contrast). The background of each microscopic field was subtracted with the Subtract Background method of the ImageJ [45] software using a rolling ball radius of 300 [46]. Before analyzing we manually stitched the images together using the Photoshop 7.0 software and smoothed them running 3 times the Smooth algorithm of the ImageJ software to decrease noise. This filter replaces each pixel with the average of its 3x3 neighborhood. After stitching the whole field of view was as large as 13000*10000 pixels (8,5*6,5 mm) containing approx. 100-150 000 astrocytes. Virus infection of single cells monitored by GFP expression develops in time with an increasing fluorescent intensity. Depending on the initial expression level of infected cells and the sensitivity of the detection system an intensity threshold has to be declared in order to decide whether a cell is already infected. On the basis of visual image analysis we tuned the threshold to find all infected cells even in a very early state of infection appearing brighter than the background noise. As a result we gained a binary black and white image. (See **Fig. 3.**)

**Real time monitoring of infection spreading.**

Dynamics of virus propagation was monitored in astroglial cultures treated with different titers of BDG virus. GFP expression within infected cells allowed the real time monitoring of the spread of infection. Time-lapse recordings were performed on a computer-controlled Leica DM IRB inverted microscope equipped with 10x and 20x objectives and an Olympus DP70 cooled CCD camera. We used Märzhäuser Scan IM 120x100 motorized stage and fine focus motor. A maximum number of 4 cell cultures in 35 mm Greiner or Ibidi Petri dishes were kept at 37°C in a humidified 5% $CO_2$ atmosphere within a CellMovie microscope stage incubator. For time-lapse imaging we used the CellMovie control software (http://cellmovie.eu). Phase-contrast and fluorescent images were acquired every 10-20 min for 2–5 days. We scanned 3x3 or 4x4 adjoining fields of view of cell cultures in every cycle of the time-lapse program. Automated focus correction was carried out in phase contrast mode in each time-lapse cycle.

**Analysis of virus propagation dynamics.**

The image processing of the movies was carried out in a partially different way. After the experiment the background of each microscopic field was subtracted with the Subtract Background method of the ImageJ software using a rolling ball radius of 300. To gain mosaic-like large scale jpeg images from the 3x3 or 4x4 microscopic fields we stitched them on the basis of the *xy* coordinates of the motorized microscope stage. Stitched jpeg images were also converted to mpeg videos with the help of a Linux script using the mjpegtools graphical software package. See supplementary **Videos S1, S2,** and **S3.** The dynamics of virus spreading is considered to be unaffected by the intensity threshold described above if we apply a constant value for all the consecutive time-lapse images: increasing or decreasing the threshold in the range of early infection results only in a time shift of virus spreading. To test whether the intensity threshold influences our result**s** we applied different values and found the same behaviour of graphs. Microscopic fields in each culture were than analyzed with the WTrack software developed by B. Gönci (unpublished). The geometric centres of GFP+ cells were tagged manually one by one on the basis of visual detection resulting in the xy position of infected cells in each time-lapse frame during the first 24 ("high" titer) or 30 hours ("low" titer). To get maximal precision we used Wacom Intuos Pen Tablet for cell tracking. The number of infected cells grew in a quasi-exponential manner. These data were used in the development of the virus propagation model. See supplementary **Videos S4, S5, and S6.**

**Simulations, modeling.**

The model we designed to simulate infection spreading is based on the combination of the well known concepts of percolation and the SI or SIS (susceptible-infected-susceptible) type epidemiological models [1-2.] or, in other words, it is a simple infection spreading model in an inhomogeneous environment with a simple assumption about the distribution of the susceptibilities of the individual cells. Our goal was to reproduce the dynamics of infection spreading and the structure and shape of the resulting clusters of infected cells as observed in the experiments using a model as simple (thus, as universal) as possible. The simulations have been carried out on a hexagonal lattice to minimize the effect of lattice geometry-dependent, preferred orientations. The time during the simulation proceeds in discrete steps. In our case a single step corresponds to one minute in the experiment.

Virus spreading was modeled by Monte Carlo simulation.

1. There are two stages when infected cells appear because of the infection from the surrounding medium. At the beginning of the experiment the whole system is exposed to the infection uniformly, thus, infected cells appear randomly in time and space according to a rate taken as input from the experiments. The number of cells became infected due to the initial exposure to the infection was approximated from **Fig. 8.B**: the frequency with which the randomly positioned infected cells appear during this stage can be well approximated with a lognormal distribution. This is the dominant mechanism for infection from the medium. At later stages cells (primarily during lysis) release virus particles into the medium resulting in a constantly increasing, but relatively low titer, leading to a second stage of infection beginning at around 18 hours (See also the increase in the number of appearing clusters at around 18 hours in **Fig. 8.**). This stage was approximated by a linearly increasing probability that new, infected cells will appear, starting at 20 hours, and lasting until 42 hours, when the simulation was stopped, because by this time the death rate of cells becomes non-negligible. To account for the death of the cells a more complicated model would be needed, and is out of the scope of this study.

2. In every simulation step, the yet healthy immediate neighbours of the set of already infected cells are considered for possible infection. Cell A infects its not yet infected neighbour B in a certain cycle of the simulation if a random number r chosen from the [0,1] interval in this cycle is smaller than the $P_i$ susceptibility value of cell B. Susceptibility value of each cell is chosen randomly and fixed at the beginning of the simulation: $P_i = \rho p_i^\alpha$. Here $p_i$ is a random number generated from a random variable X having continuous uniform distribution on the [0,1] interval. The random variable $Y = X^\alpha$ will have a probability density function of

$$\mathrm{pdf}(y) = \frac{1}{\alpha} y^{1/\alpha - 1}$$

on the [0,1] interval, zero everywhere else. This distribution infers that the low value of susceptibility is more frequent than the high value if α>1. Either power law or lognormal distribution, the final results are very similar. The constant $\rho$ was chosen (fitted to the observations) in such a way that the largest cluster of infected cells percolated at about the same time as in the experiments. (see supplementary **Fig. S4** and **Video S7**). The picturesque meaning is $\sigma = dt/T$, where dt is the time step of the simulation, T is the time needed to infect the most susceptible cells having an infected neighbour. Parameter α was tested for different integer values from 0 to 5, and the one for which the results had the best overall agreement with the experiment was used in the final model (α=3).

To minimize the effect of the system's finite size, the number of particles was double of the number of cells' appearing on the stitched experimental image [Fig 3]. The fractal dimension (**Fig. 5.**) and cluster size distribution (**Fig. 4. and Fig. 6**.) are averaged results of 10 distinct simulations, done with the same parameters, in order to avoid accidental good/bad agreement with the experiment due to the variance of the simulation results. The parameters of the model are summarized in **Table 1.**

| Parameter type | Value | Mode of determination |
|---|---|---|
| System size | 500x500 particles in a hexagonal lattice | Double of the size of the partial system captured by the images |
| Number of particles appearing during the first stage of infection | 500 | Approximated from **Fig. 8.** |
| Number of particles appearing in the second stage | 2000 | Best fit |
| $\rho$ | 0.0182 | Best fit |
| Power law exponent: $\alpha$ | 3 | Best fit |
| Expected value of the lognormal distribution (used for the first phase of infection) | 7.5 hours | Fitted to experimental results (see **Fig. 8.**) |
| Standard deviation of the lognormal distribution (used for the first phase of infection) | 2.5 hours | Fitted to experimental results (see **Fig. 8.**) |

*Table 1.*
*The main parameters of the simulation. Best fit means that the parameter was optimized to give best overall agreement with the experimental results.*

Unfortunately, standard tests for comparing the very complex experimental and computer generated fractal-like clusters are not available. Thus, to obtain results supporting the validity of our model, we have decided to compare the experimental results with two possible models, one being the model we propose, the other one being a standard percolation model, and evaluate the differences between the numbers we calculated from the observations with those obtained from the simulations. When applying the standard model, we assumed the same specific dynamics for the spatial and temporal dependence of the new centres of infection. However - instead of the non-linear function describing the probability that an infected cell spreads the viruses to its neighbours - we used a common and simpler assumption of this probability being a constant p=0.7. In addition we have also visually compared the actual appearance of the cluster shapes. The results of these studies can be summarized as follows:

The best fit to the fractal dimension of the largest cluster in **Fig. 5.** results in a value D=1.78±0.03. Our non-linear model gives a result D=1.813±0.027. The standard percolation model leads to the value D=1.861 ± 0.019. Thus, we conclude that regarding our model versus the experiment, the error bars overlap (there is a reasonable agreement), while a standard percolation model gives a result which is outside of the region allowed by the error bars. In the language of the t-test, the agreement of our model is still not good enough to establish a 95% confidence of its validity, but the probability that our model is superior to the simpler standard version is orders of magnitude higher. The situation is very similar in the case of the exponent describing the cluster size distribution. Here the experimentally obtained value of -1.7±0.04 can be compared with the values -1,643±0,03 (our model) and

-1,578±0,02 (standard model). Again, the results are such that there is an overlap within the error bars for our model and there is no overlap for the standard model. **Fig. S5** shows the visual result gained using the standard percolation model.

Visual observation of the experimental and the simulated clusters suggests that the relatively smooth and rounded hull of the largest cluster for the standard model (p=0.7) would result in a fractal line with a significantly smaller dimension and a much smaller scaling region (length versus overall extension) than hulls of the experimental and the simulated (our model) clusters. Indeed, the numbers we obtain are for the fractal dimensions of the hulls, $D_H$, $D_H$(experiment)=1.54±0.06, $D_H$(our model)=1.57±0.06 and $D_H$(standard model)=1.41±0.06.

**Analysis of the clusters.**

The resulting configurations (clusters of connected infected cells) from the simulation and also from the experiment have been analyzed by methods of statistical physics. The cluster size distribution has been determined using the Hoshen-Kopelman algorithm to find clusters [47]. The fractal dimension of the biggest cluster and the average of the fractal dimensions of all clusters above a given size were calculated from density correlations [25]. In the case of the experimental results the pixels of the pictures belonging to a cluster were considered as particles. (see **Fig.4. b,d,f** and **Fig. S.2.**)

**Percolation theory essentials.**

One of the classic explanatory examples for standard percolation processes involves a forest, in which a few trees spontaneously become infected. The probability that a neighbouring tree becomes infected by an already infected one depends on several factors and is modelled with allowing infection between two trees with a given probability p. In the case of a tree nursery the trees would be, for example, arranged as the nodes of a square lattice. According to percolation theory if p is smaller than a given, lattice dependent $p_c$, the infection cannot spread, above this critical probability a finite proportion of the trees become infected, while at the critical point the infected trees form a large very irregular "cluster" of neighbouring trees. This cluster can be described in terms of fractal geometry, because the number of trees N(L) in a region of linear size L scales as $N(L) \propto L^D$, i.e., instead of a power equal to 2 (corresponding to a 2 dimensional cluster) the fractal dimension of the cluster, D, is a number smaller than 2 [25]. Here L is measured using a unit being equal to the average distance between two neighbours or, in the case of lattice geometry, it is the lattice constant. Another helpful thought experiment for demonstrating the phenomena taking place in percolation systems involves coins dropped with some rate at randomly chosen spots of a given area. As the number of coins on the ground grows the probability that two nearby coins eventually overlap increases, thus, larger and larger clusters appear. At a critical number of the coins (corresponding to the critical probability above) the largest cluster (of overlapping coins) spreads over the whole area. Meanwhile the distribution of the cluster sizes is very uneven, there will be many small ones, a lesser number of larger ones, and a single so called giant cluster. This distribution is denoted by $n_s$ (where $n_s$ is the number of clusters consisting of s units) and is described by a power law with an exponent τ, i.e., $n_s \propto s^{-\tau}$ which is very different from the usual Gaussian distribution since it means that the cluster sizes are not scattered around a well defined average [23].

ACKNOWLEDGEMENTS

We thank Dr. Zsolt Boldogkői and Dr. Ákos Hornyák for the recombinant BDG virus. We are grateful to Emília Madarász and Krisztina Kovács for supporting our work.

SUPPORTING INFORMATION

**Figure S1.**

Original image of **Figure 3**. 10x10 microscopic field images were taken from fixed cultures with GFP expressing infected cells, 48 hours after the onset of infection with "high" virus titer. High resolution file can be downloaded from: http://amur.elte.hu/BDGVirus/

**Figure S2.**

Rate of appearance of new cluster seeds as a function of time in the model. Simulation starts with 500 seeds with a "switching on" time following a log-normal distribution fitted to the experimentally observed parameters (mean = 7.5h, standard deviation = 2.5 h), and continues with 2000 further seeds with a constantly increasing rate in time between 20 h and 42 h corresponding to the observed second wave of infection. High resolution file can be downloaded from: http://amur.elte.hu/BDGVirus/

**Figure S3.**

Features of the astrocytic cultures. Astrocytic cultures were live-stained with a fluorescent plasma membrane stain (**A**,**B**,**C**,**H**) and the nuclei were labeled with DAPI (**C**,**D**,**E**,**H**). Both the membrane-stained and the phase contrast (**F**,**G**) images show that astrocytes are arranged to a mosaic like monolayer. Scale bars: 25 μm (**A**,**D**,**F**,**H**), 10 μm (**B**,**C**,**E**,**G**). High resolution file can be downloaded from: http://amur.elte.hu/BDGVirus/

**Figure S4.**

Comparing the experimentally and computationally obtained clusters. Comparison of clusters found by the Hoshen – Koppelman algorithm on photo montages (**A**,**B**,**C**) and clusters developed in the model (**D**,**E**,**F**). Time elapsed after infection was 18h (**A**,**D**), 30h (**B**,**E**) and 42h (**C**,**F**), respectively. Cluster growth, imitating virus propagation in the model can be seen in supplementary **Video 7.** High resolution file can be downloaded from: http://amur.elte.hu/BDGVirus/

**Figure S5.**

Clusters obtained in the simulations of a simplified model. The clusters generated with seeds of infection appearing at a rate corresponding to the experiment, but the infection spreading according to the p=0.7 simple percolation rule leads to images like this one. The clusters, on the scale being about 20 to 50 lattice units (a lattice unit corresponding to the size of a cell), are much more compact than the ones observed in the experiments, and their percolation-like features start to show up only at a much larger scale than in the experiments. High resolution file can be downloaded from: http://amur.elte.hu/BDGVirus/

|                | Original [%] | 1st wash [%] | 2nd wash [%] | 3rd wash [%] |
|----------------|--------------|--------------|--------------|--------------|
| Low titer, 1 h  | 67           | 7            | 0            | 0            |
| Low titer, 24 h | 63           | 17           | 0            | 0            |
| High titer, 1 h | 100          | 10           | 7            | 0            |
| High titer, 24 h| 100          | 33           | 13           | 7            |

**Table S1.**

**Removal of infective viral particles from the culture medium.** Percentage of infected (GFP expressing) cell groups developed in astrocytic cultures incubated with the first 3 washes of virus treated "original" sister-cultures for 1h or 24 hours. "Low titer": $2{,}5\times10^4$ PFU/ml, "High titer": $2{,}5\times10^5$ PFU/ml.

| High titer CM | Original [%] | 1 h [%] | 24 h [%] |
|---------------|--------------|---------|----------|
| 1 h           | 0            | 0       | 0        |
| 5 h           | 0            | 0       | 0        |
| 24 h          | 63           | 0       | 7        |
| 30 h          | 80           | 7       | 20       |
| 48 h          | 100          | nd      | 100      |

**Table S2.**

**Release of infective viral particles to the culture medium by astrocytes.** Percentage of infected (GFP expressing) cell groups developed in astrocytic cultures incubated for 1h or 24hrs with conditioned media (CM) taken from virus treated sister-cultures. nd: not detected

**Video S1.**

Virus spreading **(**development of GFP expression) in an astroglial cell culture infected with "low" titer ($2{,}5\times10^4$ PFU/ml) BDG-PRV virus. High resolution videos can be downloaded from: http://amur.elte.hu/BDGVirus/

**Video S2.**

Virus spreading (development of GFP expression) in an astroglial cell culture infected with "high" titer ($2.5\times10^5$ PFU/ml) BDG-PRV virus. High resolution videos can be downloaded from: http://amur.elte.hu/BDGVirus/

**Video S3.**

Virus spreading (development of GFP expression) in an astroglial cell culture infected with "highest" titer ($1{,}25\times10^6$ PFU/ml) BDG-PRV virus. High resolution videos can be downloaded from: http://amur.elte.hu/BDGVirus/

**Video S4.**

The video shows the appearance of GFP in infected cells over time within a representative microscopic field. High resolution file can be downloaded from: http://amur.elte.hu/BDGVirus/

**Video S5.**

Graphical representation of **Video 4.** showing virus propagation within an astroglial culture treated with low titer ($2,5 \times 10^4$ PFU/ml) BDG-PRV virus. Red cells indicate the foci of the developing clusters. High resolution file can be downloaded from: http://amur.elte.hu/BDGVirus/

**Video S6.**

Graphical representation of virus propagation within 9 adjacent microscopic fields of an astroglial culture treated with low titer ($2,5 \times 10^4$ PFU/ml) BDG-PRV virus. Graphical representations of virus propagation within an astroglial culture treated with low titer ($2,5 \times 10^4$ PFU/ml) BDG-PRV virus. High resolution file can be downloaded from: http://amur.elte.hu/BDGVirus/

**Video S7.**

Cluster growth, imitating virus propagation in the model. High resolution file can be downloaded from: http://amur.elte.hu/BDGVirus/